\documentclass[%
 reprint,
 amsmath,amssymb,
 aps,
 pra,
]{revtex4-1}

\usepackage{graphicx}
\usepackage{dcolumn}
\usepackage{bm}
\usepackage{mathtools}

\begin{document}


\title{The structure of quadratic Gauss sums in Talbot effect}

\author{Carlos R. Fern\'andez-Pousa}
\email{c.pousa@umh.es}
\affiliation{Department of Communications Engineering, Universidad Miguel Hern\'andez, Av. Universidad s/n, E03203 Elche, Spain}
\date{November 29, 2016}

\begin{abstract}
The field diffracted from a one-dimensional, coherently illuminated periodic structure at fractional Talbot distances can be described as a coherent sum of shifted units cells weighted by a set of phases given by quadratic Gauss sums. We report on the computation of these sums by use of the properties of a recently introduced integer $s$, which is constructed here directly from the two coprime numbers $p$ that $q$ that define the fractional Talbot plane. Using integer $s$, the computation is reduced, up to a global phase, to the trivial completion of the exponential of the square of a sum. In addition, it is shown that the Gauss sums can be reduced to two cases, depending only on the parity of integer $q$. Explicit and simpler expressions for the two forms of integer $s$ are also provided. The Gauss sums are presented as a Discrete Fourier Transform pair between periodic sequences of length $q$ showing perfect periodic autocorrelation.
The relationship with one-dimensional multilevel phase structures is exemplified by the study of Talbot array illuminators. These results represent a simple means for the design and analysis of systems employing the fractional Talbot effect.

\begin{description}
\item[Keywords]{Talbot effect, quadratic Gauss sums, perfect periodic autocorrelation, Talbot array\\illuminators.}
\end{description}

\end{abstract}

\maketitle


\section{Introduction}

When a diffraction grating is illuminated by a coherent plane wavefront, a series of self-images can be observed at certain distances beyond the object, at integer multiples of a fundamental Talbot distance \cite{talbot1836}, \cite{rayleigh1881}. A related phenomenon occurs at fractional values $p/q$ of this distance, where the field is a coherent superposition of $q$ shifted and weighted copies of the grating's unit cell \cite{winthrop1965}. These phenomena are consequence of the interference between diffraction orders, which acquire a quadratic phase under Fresnel propagation, and have also been observed in other domains, such as the angular spectrum \cite{azana2014angular} and, through the space-time duality \cite{akhmanov1969nonstationary}, \cite{kolner1994space}, in time \cite{azana2001temporal} and in the optical spectrum \cite{azana2005spectral},  \cite{caraquitena2011spectral}. As simple diffractive effects, they have found widespread application \cite{patorski89}, \cite{wen13talbot}. 

Much of this work was triggered by the publication, twenty years ago, of a seminal paper by Berry and Klein \cite{BK96}, who showed the relationship of the weighting factors at fractional Talbot distances with the classical quadratic Gauss sums of number theory. The solution to these Gauss sums was presented in \cite{BK96} by considering three separated cases, relying on the result of a previous investigation \cite{HB80}. One of these three solutions contained an error which was corrected later by Matsutami and \^Onishi \cite{MO03}. More recently \cite{LHA15}, it has been shown that a proper rewording of the original threefold form of the Gauss sums leads to a compact description in terms of the product of two phases, one that is constant and only depends on integers $p$ and $q$ and a second that is quadratic and proportional to a new integer $s$. The observation in \cite{LHA15} can be stated as the fact that the Discrete Fourier Transform (DFT) of a quadratic phase sequence is itself a quadratic phase sequence. In addition to its intrinsic beauty, this correspondence finds immediate applications as a simple rule to implement the phases in physical systems employing fractional Talbot effect. However, the characterization of the integer $s$ that arises in the Gauss sums was obtained in \cite{LHA15} by direct comparison, relying on the original computations in \cite{HB80} and \cite{MO03}. 

In this paper, we revisit the computation of these Gauss sums from a new perspective. Using standard techniques from number theory, integer $s$ is constructed in an abstract way from the two coprime integers $p$ and $q$ that define the fractional Talbot effect. The properties shown by this integer allows for an almost trivial derivation of the main property of the Gauss sum, namely its quadratic dependence. It is also shown that both $s$ and the result of the Gauss sums can be expressed only in two cases associated to the parity of integer $q$ and, in addition, that both members of the DFT pair that describe the Gauss sums show perfect periodic autocorrelation \cite{luke1988sequences}, \cite{chu1972polyphase}. These results not only simplify the description of self-imaging phenomena, but also permit a compact analysis of systems employing the fractional Talbot effect. As an example, the relationship with multilevel phase structures is exemplified by the detailed study of one-dimensional Talbot array illuminators (TAI) \cite{lohmann1990making},
\cite{leger1990efficient},
\cite{arrizon1993talbot},
\cite{szwaykowski1993talbot},
\cite{arrizon1994multilevel},
\cite{zhou1999analytic}.

In our presentation we will employ the description of Talbot effect in the time domain \cite{azana2001temporal}, where temporally modulated periodic waves substitute for diffraction gratings and dispersion for diffraction. Although our results can be extended to different domains as pointed out before, the use of the temporal formalism is preferred here because it conforms to the standard definition of discrete-time transforms of signal theory, which is the natural language upon which the main results are presented. 

The plan of the paper is the following. In Section II we review the basic result of \cite{BK96} about the structure of Talbot self-images of periodic objects. In Section III, we construct the integer $s$ and derive several properties, and in Section IV we compute the Gauss sums of Talbot effect and derive their autocorrelation properties. Section V is devoted to the analysis of TAIs and Section VI presents our conclusions. The paper is completed with three appendices which present some results and definitions from number theory and two lengthy computations.

\section{The temporal Talbot effect}

Let us consider a modulated optical wave, whose electric field is given by $\mathcal{E}(t) = E(t) \exp(j\omega_0t)$, with $\omega_0$ the frequency of the optical carrier and $e(t)$ the optical complex envelope or baseband representation of the wave. The envelope is assumed periodic with fundamental period $T$, and represented as the repetition of function $w(t)$:
\begin{equation}\label{input}
E(t)=\sum_{n=-\infty}^{+\infty}\delta\left(t-nT\right)\otimes w(t),
\end{equation}
where $\otimes$ denotes convolution. Here we assume that $w(t)$ is contained in a period, $0\leq w(t)< T$, and so it describes the unit cell of the periodic train. The spectrum of the optical envelope $E(t)$ is contained in a set of equally-spaced angular frequencies, $\omega_n = 2\pi n/T$ with $n$ integer. 

This field feeds a linear delay line with lowest-order dispersion. Without loss of generality, the transfer function that describes the linear propagation of the envelope $E(t)$ can be assumed quadratic, $H(\omega)=\exp(-j\phi\omega^2/2)$, where $\phi$ is the dispersion coefficient, equal to the derivative of the group delay with angular frequency at the carrier value, $\omega_0$. The temporal Talbot effect manifests itself as the recovery of the initial train, or of a periodic structure directly related to the original train, at certain values of dispersion $\phi$ characterized by the period $T$ and two positive and mutually coprime integers $p$ and $q$. The defining equation is: 
 \begin{equation}\label{talbot}
2\pi |\phi| = \frac{p}{q} T^2.
\end{equation}
At these values, $H(\omega_n)=\exp(-j\pi{\sigma}_{\phi}\frac{p}{q}n^2)$, where ${\sigma}_{\phi}$ is the sign of $\phi$. The output envelope $E'(t)$ can be computed in the spectral domain by use of the Poisson formula in (\ref{input}) and the values of $H(\omega_n)$,
\begin{align}
E'(t)=\frac{1}{T}\left(\sum_{n=-\infty}^{+\infty}
e^{-j\pi \sigma_\phi\frac{p}{q}n^2}e^{j2\pi n\frac{t}{T}}
\right)\otimes w(t).
\end{align}
Note that the value of the sum for $\sigma_\phi=-1$ is the complex conjugated to the sum for $\sigma_\phi=1$, since the change in the sign of the exponential can be compensated with an inversion $n\rightarrow -n$. This first part of the equation can be decomposed by changing the sum over $n$ to a sum over the pair of integers $\ell=0,\dots, q-1$ and $m=-\infty,\dots,+\infty$, with $n=\ell + mq$. The sum over $m$ can be performed using again the Poisson formula, and we get \cite{BK96}:
\begin{align}\label{tte}
E'(t)&=\delta\left(t-e_{pq} \frac{T}{2}\right)
\otimes\sum_{k=-\infty}^{+\infty}\delta\left(t-kT\right)\nonumber\\
&\otimes\frac{1}{\sqrt{q}}\sum_{n=0}^{q-1} e^{j\sigma_\phi\xi_n}\delta\left(t-n\frac{T}{q}\right)\otimes w(t),
\end{align}
where $e_{x}$ represents the parity of integer $x$, so that $e_{x}=0$ when $x$ is even and $e_{x}=1$ when $x$ is odd. In (\ref{tte}), $e_{pq}$ thus represents the parity of the product of integers $pq$. The phases in (\ref{tte}) are given by the following Gauss sums:
\begin{align}\label{gauss}
e^{j\sigma_\phi\xi_n} &=
\frac{1}{\sqrt{q}} \sum_{m=0}^{q-1} \, e^{-\pi j\sigma_\phi \frac{p}{q}m^2}(-1)^{pqm} e^{2\pi j \frac{nm}{q}}\nonumber \\
&= \frac{1}{\sqrt{q}} \sum_{m=0}^{q-1} \, e^{-\pi j\sigma_\phi\frac{p}{q}(1+qe_q)m^2} e^{2\pi j \frac{nm}{q}}.
\end{align}
In the final part of the equation we have used that $(-1)^{pqm}=(-1)^{pe_qm^2\sigma_\phi}$. This second form of the Gauss sums will be the starting point of our computation in Section IV. 

According to (\ref{tte}), the output is composed of a repetitive structure of the same period $T$ as the original train, explicitly recognized in the second term of the multiple convolution. This train is shifted by half a period if $pq$ is odd, as shown by the first delta term. The repetitive structure that defines the output train is described by the two last terms in (\ref{tte}), and is composed of the coherent sum of $q$ replicas of the input unit cell, $w(t)$, mutually shifted by $T/q$ and weighted by factors $\exp(j\sigma_\phi\xi_n)/\sqrt{q}$. We point out that, expect in the case that $w(t)$ is contained in an interval of length $T/q$, the last line in (\ref{tte}) does not describe the unit cell of the output, since this expression may extend beyond an interval of length $T$.

When $q=1$, the output train is similar to the original. The resulting wave is referred to as an integer Talbot image of order or index $p$, or simply a self-image of order $p$ of the input train. If, in addition, $p$ is odd, it is usually referred to as a shifted integer Talbot image. In general, the resulting trains for $q > 1$ are called fractional Talbot images of order $p/q$, and shifted when the product $pq$ is odd. To proceed with our analysis we devote the following sections to the characterization of phases $\exp(j\xi_n)$ in terms of an integer $s$ constructed from $p$ and $q$.

\section{A parity-dependent modular inverse}

The main results concerning the properties of integer $s$ is stated in the following theorem:\\

\textbf{Theorem 1.} Given two coprime and positive integers, $p$ and $q$, there exists a unique integer $s$ such that it verifies the following three properties:

(a) $s$ lies in the range $1\leq s \leq 2q-1$,

(b) $s$ is a solution of the modular equation:
\begin{equation}\label{lmdc}
sp=1+qe_{q}\ ({\rm mod}\ 2q), 
\end{equation}

(c) $s$ has opposite parity to $q$.

{\noindent In addition, such a integer $s$ also verifies:}

(d) $s$ is coprime with $q$. \\

The theorem will be proved in four steps. First, we show the existence of integers verifying (a) and (b) by the explicit construction of solutions to (\ref{lmdc}). Second, we will show that these solutions also verify (c). Third, we show that any other solution of (\ref{lmdc}) is either out of the range (a) or does not verify (c). Finally, we deduce that $s$ and $q$ must be coprime. For the definition of the symbols used in the proof we refer to Appendix \ref{appnt}.

\textbf{Existence.} For even $q$, Eq. (\ref{lmdc}) reduces to $sp=1$ (mod $2q$), and since integers $p$ and $q$ are mutually prime, so are $p$ and $2q$. Then the equation can solved as:
\begin{equation}
s=\left[\dfrac{1}{p}\right]_{2q}\ \   \ \ \ (q         \ \rm{even}),
\label{s-qpar}
\end{equation}
where $[1/a]_n$ is the inverse of $a$ (mod $n$). In particular, the inverse of $p$ (mod $2q$) lies in the range $1\leq s\leq 2q-1$. 

When $q$ is odd, integers $2p$ and $q$ are coprime, and a solution to (\ref{lmdc}) is given by:
\begin{equation}
s=2\left[\dfrac{1}{2p}\right]_{q}\ \   \  \ \ (q         \ \rm{odd}).
\label{s-qimpar}
\end{equation}
Indeed, in this case the product $sp$ is: 
\begin{equation}\label{t}
sp=2p\left[\dfrac{1}{2p}\right]_{q}=1\ ({\rm mod}\ q)= 1+aq,
\end{equation}
for a certain integer $a$. But this $a$ must be odd, since $s$ in (\ref{s-qimpar}) is even and $q$ is odd, and therefore (\ref{t}) can be written as: 
\begin{equation}
sp=1+q\ ({\rm mod}\ 2q),
\end{equation}
thus solving equation (\ref{lmdc}). Now, since the inverse of $2p$ (mod $q$) is contained in the interval $1\leq [1/2p]_q\leq q-1$, integer $s$ lies in the range stated in the theorem. 

\textbf{Parity.} As for property (c), the statement is trivial for the solution found for $q$ odd, since the explicit construction (\ref{s-qimpar}) shows that in this case $s$ is even. For $q$ even we first note that $p$ must be odd. Now, the solution of (\ref{lmdc}) is of the form:
\begin{equation}
sp = 1 + 2bq
\end{equation}
for certain integer $b$. The right hand side of this equation is odd, and therefore $s$ must be odd. Note that the explicit solutions (\ref{s-qpar}) and (\ref{s-qimpar}) together with property (c) point out that $s$ is an modular inverse of $p$ whose concrete form depends on the parity of $q$.

\textbf{Uniqueness.} Integers (\ref{s-qpar}) and (\ref{s-qimpar}) are the unique solutions to (\ref{lmdc}) in the range $1\leq s\leq 2q-1$ with opposite parity to $q$. In fact, given two solutions of (\ref{lmdc}) for given $p$ and $q$, namely $s$ given by (\ref{s-qpar}) or (\ref{s-qimpar}), and a different solution $\tilde{s}$, their difference verifies $(\tilde{s}-s)p=0$ (mod $2q$). Here we need to analyze separately the cases.

For $q$ even, and since $p$ and $2q$ are coprime, this implies that $\tilde{s}-s =0$ (mod $2q$) and so the uniqueness of $s$ in the range $1\leq s\leq 2q-1$ follows. The same argument applies for $q$ odd if we further assume that $p$ is also odd, since again $p$ and $2q$ are coprime.

The only remaining case is $q$ odd and $p$ even. The equation above, $(\tilde{s}-s)p=0$ (mod $2q$), only implies that $(\tilde{s} -s)=0$ (mod $q$), and so both (\ref{s-qimpar}) and $\tilde{s}=s+q$ are solutions to (\ref{lmdc}), as it canbe easily checked. In particular, $\tilde{s}$ may lie in the range $0\leq \tilde{s}\leq 2q-1$. For example, for $p=2$ and $q=3$, both $s=2$ and $\tilde{s} =5$ are solutions of (\ref{lmdc}) in the range $1\leq s, \tilde{s} \leq 5$. However, integer $s$ as given by (\ref{s-qimpar}) is selected by the fact that it has the opposite parity to $q$, contrary to $\tilde{s}=s+q$. This completes the proof.

\textbf{Coprime.} Finally, to show that integer $s$ is indeed coprime with $q$, let us suppose that $s$ and $q$ share a common factor, say $c$. We will show that $c= \pm 1$. First, note that $c$ is also a factor or the product $sp$. Then, dividing (\ref{lmdc}) by this factor we get:
\begin{equation}
\frac{sp}{c}=\frac{1}{c}+
\frac{qe_q}{c}\  \  \ \left({\rm mod}\ \frac{2q}{c}\right).
\label{coprime}
\end{equation}
Now, since the left hand side of this equation is an integer and $c$ is a factor of $q$, then $1/c$ should be integer, and thus $c=\pm 1$. The proof of Theorem 1 is complete.\\

We point out that the expressions for $s$ derived in \cite{MO03} and used in \cite{LHA15} are different from the ones given by (\ref{s-qpar}) and (\ref{s-qimpar}). The equivalence is shown in Appendix \ref{appeq}.

\begin{table}[t]
\caption{\label{tabla}Values of integer $s$ in the range $0\leq s\leq 2q-1$ as a function of $p$ and $q$.}
\begin{ruledtabular}
\begin{tabular}{ c c | c c c c c c c c c c c c}
   &   &   &   &   &   &\textbf{\textit{p}}&   &   &   &   &    \\
   &   & \textbf{1} & \textbf{2} & \textbf{3} &\textbf{ 4} & \textbf{5} & \textbf{6} & \textbf{7} & \textbf{8} & \textbf{9} & \textbf{10} \\ \hline
   & \textbf{2} & 1 &   & 3 &   & 1 &   & 3 &   & 1 &    \\
   & \textbf{3} & 4 & 2 &   & 4 & 2 &   & 4 & 2 &   & 4  \\
   & \textbf{4} & 1 &   & 3 &   & 5 &   & 7 &   & 1 &    \\
   & \textbf{5} & 6 & 8 & 2 & 4 &   & 6 & 8 & 2 & 4 &    \\
\textbf{\textit{q}}
   & \textbf{6} & 1 &   &   &   & 5 &   & 7 &   &   &    \\
   & \textbf{7} & 8 & 4 & 12& 2 & 10& 6 &   & 8 & 4 & 12 \\
   & \textbf{8} & 1 &   & 11&   & 13&   & 7 &   & 9 &    \\
   & \textbf{9} & 10& 14&   & 16& 2 &   & 4 & 8 &   & 10 \\
   & \textbf{10}& 1 &   & 7 &   &   &   & 3 &   & 9 &    \\
   & \textbf{11}& 12& 6 & 4 & 14& 20& 2 & 8 & 18& 16& 10 \\
\end{tabular}
\end{ruledtabular}
\end{table}

The computation of the values of $s$, as a function of $p$ and $q$, can proceed from the systematic solution of (\ref{s-qpar}) and (\ref{s-qimpar}) or from an equivalent expression such as those in Appendix \ref{appeq}. A table for the lowest values of $p$ and $q$ was presented in \cite{LHA15} and reproduced in Table \ref{tabla} (see, however, note \footnote{The first row ($q$=2) in the table of \citep{LHA15} is in error. It reads 1-1-1-1-1 but it should read 1-3-1-3-1.}). The calculus of $s$ is facilitated by the observation that, regardless the parity of a given $q$, (\ref{lmdc}) implies that integer $s$ for given coprime numbers $p$ and $q$ is the same that for $p+2q$ and $q$. This property reflects the exact periodicity, without half-interval shifts, of the temporal Talbot effect. In the case of $q$ odd, it also follows from (\ref{lmdc}) that the integers $s$ for the pair $p, q$ and for the pair $p+q$ and $q$, coincide. Unfortunately, the result does not hold for $q$ even. Specific solutions can be found for several series of integers, as shown by the following result.\\

\textbf{Proposition 2.} 

(a) for any $q$: $p=1\ ({\rm mod}\ 2q) \Rightarrow s=1+qe_{q}$.

(b) for any $q$: $p=q\pm 1\ ({\rm mod}\ 2q) \Rightarrow s=q\pm 1$.

(c) for any $n, p$: $q=\pm 1 +2np \Rightarrow s=\pm 2n\ ({\rm mod}\ 2q)$.\\

\textbf{Proof.} First we note that in these three series integers $p$ and $q$ are indeed coprime, since in all cases they are defined by a relation of the form $ap+bq = \pm1$ for certain integers $a$ and $b$, and thus any common factor between $p$ and $q$ must be equal to one following a similar reasoning to that in (\ref{coprime}). Second, it is a simple observation that the parity of $s$ is the opposite to that of $q$ in all cases. Moreover, all proposed values of $s$ lie in the range $1 \leq s\leq 2q-1$, except in case (c) for which the residue mod $2q$ has been taken. Then, it suffices to check that the values of $s$ verify (\ref{lmdc}) by direct substitution. The result is trivial for (a), whereas for (b) and (c) it can be deduced from the observation that (\ref{lmdc}) is equivalent to any of the following equations:
\begin{equation}\label{sd}
sp=(q\pm 1)^2 \ \ ({\rm mod} \ 2q).
\end{equation}
In particular, (\ref{sd}) defines the {\it self-dual} integers $s=p=q\pm 1$ for which the values of $p$ and $s$ coincide. This completes the proof.\\

The calculus of $s$ is also simplified by the following result, that relates its value for {\it complementary} Talbot lines of order $p/q$ and $(q-p)/q$. Essentially it is a sum rule that applies to pairs of integers in each row of Table \ref{tabla}: \\

\textbf{Proposition 3.} If $s$ is the integer corresponding to $p$ and $q$, with $p<q$, then the integer $s'$ corresponding to $p'=q-p$ and $q$, verifies;

(a) $s'=2q-s$ for $q$ odd,

(b) $s'=q-s$ for $q$ even and $1\leq s \leq q-1$, and

(c) $s'=3q-s$ for $q$ even and $q+1\leq s\leq 2q-1$.\\

\textbf{Proof.} First notice that, if $p$ is coprime with $q$, then $p'=q-p$ is also coprime with $q$. Moreover, integer $s'$ as stated above has the same parity as $s$, so condition (c) in Theorem 1 is fulfilled. As for condition (a) in that Theorem, for $q$ odd $s'+s= 2q$, and therefore if $s$ is in the range $1\le s\leq 2q-1$ so is $s'$. For $q$ even, both (b) and (c) above gives $s'$ in the range $1 \leq s' \leq 2q-1$. Note that $s$ cannot coincide with $q$ since they must be coprime.

To complete the proof it suffices to check (\ref{lmdc}). Direct computation gives:
\begin{equation}\label{comp1}
s'p'=sp+\left[n(q-p)-s\right]q,
\end{equation}
with $n=1,2,3$ according to the statement of the proposition. When $q$ even, $s$, $p$ and $n(p-q)$ are odd, and thus $s'p'=sp$ (mod $2q$). For $q$ odd, $s$ and $n(p-q)$ are even, and again $s'p'=sp$ (mod $2q$). Therefore, since by hypothesis (\ref{lmdc}) is verified by $sp$, it is also verified by $s'p'$. The proof is complete.\\

\section{The Gauss sums}

Integer $s$ constructed in the previous section provides a simple route to evaluate the Gauss sums (\ref{tte}). Our computation is based on the observation that (\ref{gauss}) involves the expression:
\begin{equation}\label{observ}
\exp\left(-\pi j\sigma_\phi\frac{p}{q}(1+qe_q)m^2\right)
\end{equation}
which can be considered as a $q$-sequence since, for $q$ even and odd, is periodic in the running index $m$ with period $q$.  We recall that, given a sequence of $q$ complex numbers, $x_n$ with $n=0, \dots, q-1$, the Discrete Fourier Transform (DFT), denoted by $\mathcal{F}$, is defined as the $q$-sequence $X_m$ given by:
\begin{equation}\label{DFT}
X_m = \mathcal{F}(\{x_n\})_m=\sum_{n=0}^{q-1} x_n e^{-2\pi j nm /q}.
\end{equation}
The IDFT is the inverse transform to (\ref{DFT}), defined as:
\begin{equation}
x_n = \mathcal{F}^{-1}(\{X_m\})_n=\frac{1}{q} \sum_{m=0}^{q-1} X_m e^{2\pi j nm /q}.
\end{equation} 
We state the Gauss sum (\ref{gauss}) as a DFT pair:\\

\textbf{Proposition 4.} For any $p$, $q$ coprime and $s$ the integer constructed in Theorem 1, the following DFT pair holds:
\begin{align}
&x_n\equiv e^{j\sigma_\phi\xi_n} = e^{j\sigma_\phi\xi_0} \exp\left(\pi j \sigma_\phi\frac{s}{q}n^2\right) \ \ \xrightarrow{DFT} \nonumber \\
&X_m\equiv{\sqrt{q}} \exp\left[-j\pi\sigma_\phi \frac{p}{q}(1+qe_q)m^2\right],
\label{gausspair}
\end{align}
with:
\begin{align}
e^{j\xi_0} = &\left(\frac{s}{q}\right) e^{j\frac{\pi}{4}(q-1)} =\left(\frac{p}{q}\right) e^{j\frac{\pi}{4}(q-1)}\ \ (q\ {\rm odd}),\nonumber \\
e^{j\xi_0} =&
\left(\frac{q}{s}\right) e^{-j\frac{\pi}{4} s} =\left(\frac{q}{p}\right) e^{-j\frac{\pi}{4} p}\ \ (q\ {\rm even}),
\label{gaussphase}
\end{align}
and $\left(\frac{a}{b}\right)$ is the Jacobi symbol of an arbitrary integer $a$ and an odd and positive integer $b$. \\

For the definition and properties of Jacobi symbols we refer to Appendix \ref{appnt}. We also point out that the present result reduces the number of cases to two, depending only on the parity of $q$, instead to the conventional three cases studied in Refs. \cite{BK96}, \cite{HB80}, \citep{MO03} and \citep{LHA15}.\\

\textbf{Proof.} We set $\sigma_\phi =1$, the general case is considered at the end of the proof. We have to show that:
\begin{align}\label{gaussp1}
 e^{j\xi_n}=e^{j\xi_0} e^{\pi j \frac{s}{q}n^2} 
=&\frac{1}{\sqrt{q}} \sum_{m=0}^{q-1} \, e^{-\pi j \frac{p}{q}(1+qe_q)m^2} e^{2\pi j \frac{nm}{q}}.
\end{align}
First we justify the change of variables $m=sk$, with $k=0,\dots, q-1$. Since $s$ and $q$ are coprime, the values of the product $sk$ span the residues of $q$, \textit{i. e.}, the products $sk$, with $k=0,\dots, q-1$, span the set \{$0,\dots, q-1$\} mod $q$. The change $m = sk$ can thus be performed in (\ref{gaussp1}) since the contribution to the sum of a certain $m$ is invariant under shifts of the form $m\rightarrow m+q$, as it is immediate to show. With this change (\ref{gaussp1}) reads:
\begin{align}\label{gaussp2}
\frac{1}{\sqrt{q}} \sum_{k=0}^{q-1} \, e^{-\pi j \frac{p}{q}(1+qe_q)(sk)^2} e^{2\pi j \frac{s}{q}nk}.
\end{align}
We use now (\ref{lmdc}) in the first exponential of (\ref{gaussp2}): 
\begin{align}
e^{-\pi j \frac{s}{q}(1+qe_q)(sp)k^2} &=e^{-\pi j\frac{s}{q}(1+qe_q)^2k^2}\nonumber\\
&=e^{-\pi j\frac{s}{q}k^2}(-1)^{sqk}=e^{-\pi j\frac{s}{q}k^2},
\end{align}
where we have used that the product $sq$ is even. With these simplifications, the exponents in (\ref{gaussp2}) can be completed to the square of a sum, 
\begin{eqnarray}
\frac{1}{\sqrt{q}} e^{\pi j\frac{s}{q}n^2} 
\sum_{k=0}^{q-1} \, e^{-\pi j\frac{s}{q}(k-n)^2},
\end{eqnarray}
and the proof is complete if we show that:
\begin{equation}\label{gaussp3}
e^{j\xi_0} =\frac{1}{\sqrt{q}} 
\sum_{m=0}^{q-1} \, e^{-\pi j \frac{s}{q}m^2}.
\end{equation}
The fact that the sum in the right hand side of (\ref{gaussp3}) is indeed a phase can be shown by changing variables in the double sum defining its modulus squared, following a similar argument to that of an equivalent computation in \cite{BK96}. The explicit evaluation of the sum leading to (\ref{gaussphase}) can be completed by standard techniques of the same type as those used in \cite{HB80}, as is shown in Appendix \ref{appphase}. 

We observe that the quadratic phases $\exp(j\xi_n)$ are periodic and symmetric, 
\begin{equation}\label{properties}
\exp(j\xi_{n})=\exp(j\xi_{n+q})=\exp(j\xi_{-n})=\exp(j\xi_{q-n}).
\end{equation}
Now, using the general relationship for the DFT of the complex-conjugated sequence,
\begin{equation}
\mathcal{F}(\{x_n^*\})_m=\mathcal{F}(\{x_n\})_{q-m}^*,
\end{equation}
it follows from the symmetry property that the DFT of the complex conjugated sequence, $\exp(-j\xi_n)$ is the complex conjugated DFT, as stated in (\ref{gausspair}). The proof is complete.\\

We now compute the periodic autocorrelation of the $q$-sequence $x_k=\exp(j\xi_{k})$, 
\begin{equation}
R_\xi(n)=\sum_{k=0}^{q-1} \exp(-j\xi_{k})\exp(j\xi_{k+n}). 
\end{equation}
Using (\ref{gausspair}), this gives a sum of the form:
\begin{align}\label{key}
R_\xi(n)=e^{j\pi\frac{s}{q} n^2}\sum_{k=0}^{q-1}
e^{-j2\pi\frac{s}{q}kn},
\end{align}
which can be evaluated in terms of Kronecker's delta using that, for any integer $a$:
\begin{align}\label{tecn}
\sum_{k=0}^{q-1}e^{-j2\pi a k/q} =q \delta_{a,0\ ({\rm mod}\ q)}.
\end{align}
We observe from (\ref{key}) that the sum is nonzero for integers $n$ which are solutions of $sn = 0 \ ({\rm mod}\ q)$. The solutions are $n=0$ (mod $q$), since $s$ and $q$ are coprime, so:
\begin{align}\label{auto}
R_\xi(n)=q \delta_{n,0\ ({\rm mod}\, q)}.
\end{align}
The phase sequences $x_k$ shows {\it perfect} periodic autocorrelation \cite{luke1988sequences}, and therefore its spectrum, as given by (\ref{gausspair}), has magnitude $\sqrt{q}$. In fact, it can be shown that {\it both} $q$-sequences, $x_n$ and $X_m$ in (\ref{gausspair}), have perfect periodic autocorrelation, since both are examples of Chu's construction \cite{chu1972polyphase}. In that paper it was shown that, given $N$ and $M$ coprime integers, the $M$-sequences 
\begin{align}\label{chu}
\exp(\pm j\pi Nm^2/M)&\ \ \ (M\ \mathrm{even})\nonumber\\
\exp(\pm j\pi Nm(m+1)/M)&\ \ \ (M\ \mathrm{odd}),
\end{align}
have perfect periodic autocorrelation. For $q$ even, the fact that $x_k=\exp(j\xi_{k})$ belongs to this set is immediate. For $q$ odd it follows after the shift in the running index $k\rightarrow k+(q+1)/2$, since the resulting sequence is in the form (\ref{chu}) up to a global factor.

As for the Fourier-transformed sequence $X_m$, it is already in form (\ref{chu}) for $q$ even. For $q$ odd, the sequence reads;
\begin{align}
X_m= \sqrt{q} e^{-j\pi p(q+1)m^2/q}=\sqrt{q} e^{-j\pi p m^2/q}(-1)^{pm}. \end{align}
Using again the shift $m\rightarrow m+(q+1)/2$ we are led, up a global phase and the $\sqrt{q}$ factor, to:
\begin{align}
&\exp\left[-j\pi \frac{p}{q} \left(m^2+(q+1)m\right)\right](-1)^{pm}\nonumber \\
=&\exp\left[-j\pi \frac{p}{q} m\left(m+1\right)\right],
\end{align}
which is the expression in (\ref{chu}). 

\section{Talbot array illuminators}

The previous results provide a simple description of the weighted sum of unit cells that defines the fractional Talbot effect. Using (\ref{gaussp1}) in (\ref{tte}), the output field in a Talbot plane with order $p/q$ is described as the repetition of the signal:
\begin{align}\label{tproc}
\tilde w'(t)=\frac{ e^{j\sigma_\phi\xi_0}}{\sqrt{q}}\sum_{n=0}^{q-1}\exp\left(j\pi\sigma_\phi\frac{s}{q}n^2\right) w\left(t-n\frac{T}{q}\right),
\end{align}
composed of the sum of delayed input unit cells, weighted by a quadratic phase factors and the amplitude factor $1/\sqrt{q}$. In this section we use this result to analyze a concrete example.

\begin{figure}[ht]
\includegraphics[scale=0.25]{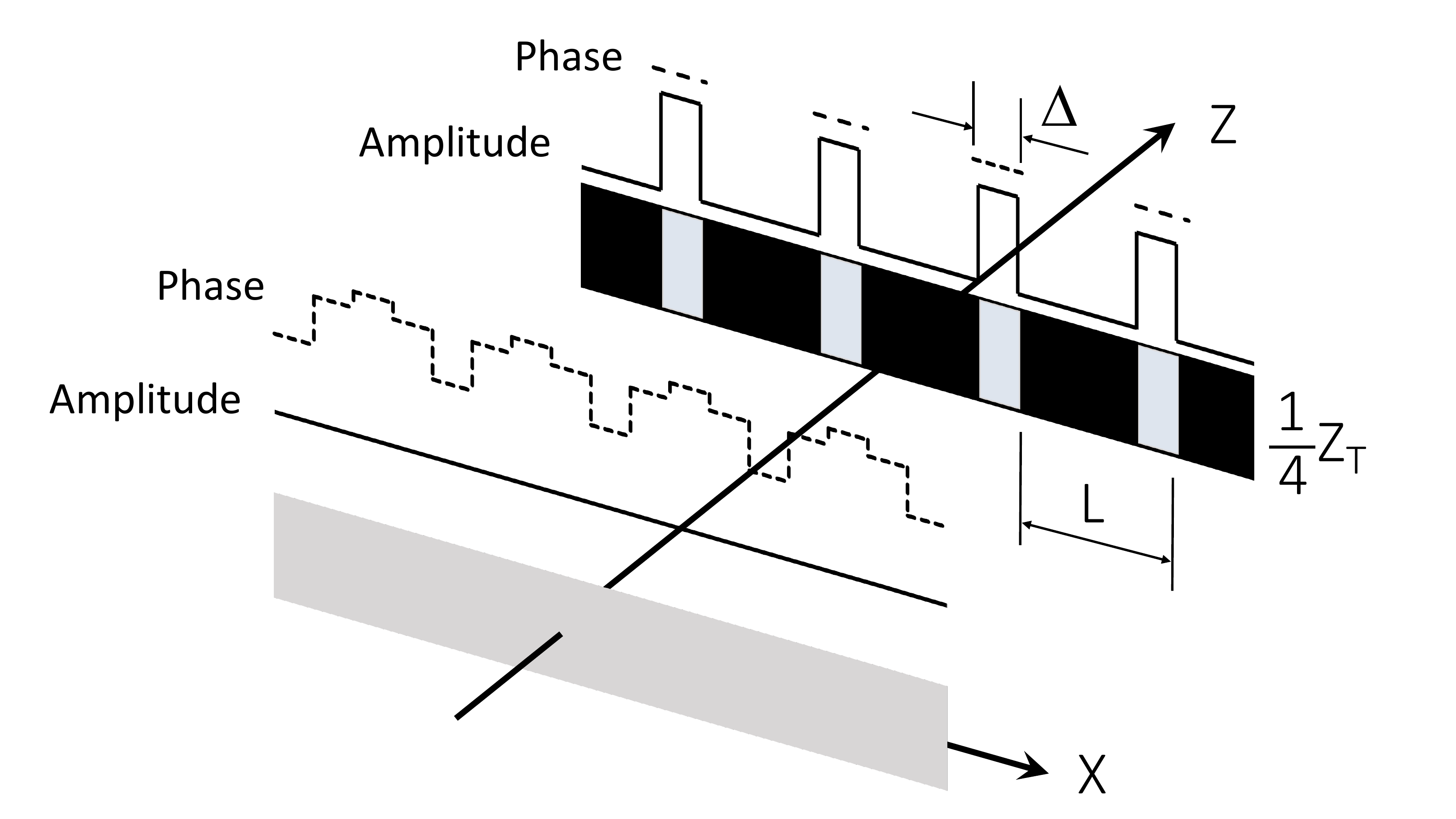}
\caption{\label{fTAI} Scheme of a Talbot array illuminator for $p/q =1/4$.}
\end{figure}

The Talbot array illuminators (TAI) are multilevel phase gratings that concentrate coherent light in a periodic array of sharp binary irradiance distributions of rectangular form, at fractional Talbot distances of order $p/q$ after the array \cite{lohmann1990making}, \cite{leger1990efficient}. Referring to Fig. \ref{fTAI}, we denote the period of the phase grating as $L$ and the width of the focusing spots as $\Delta$. The compression ratio, $\Delta/L$, is a measure of its concentration capacity, and equals the value of integer $q$ \cite{leger1990efficient}. 

\begin{figure}[ht]
\includegraphics[scale=0.25]{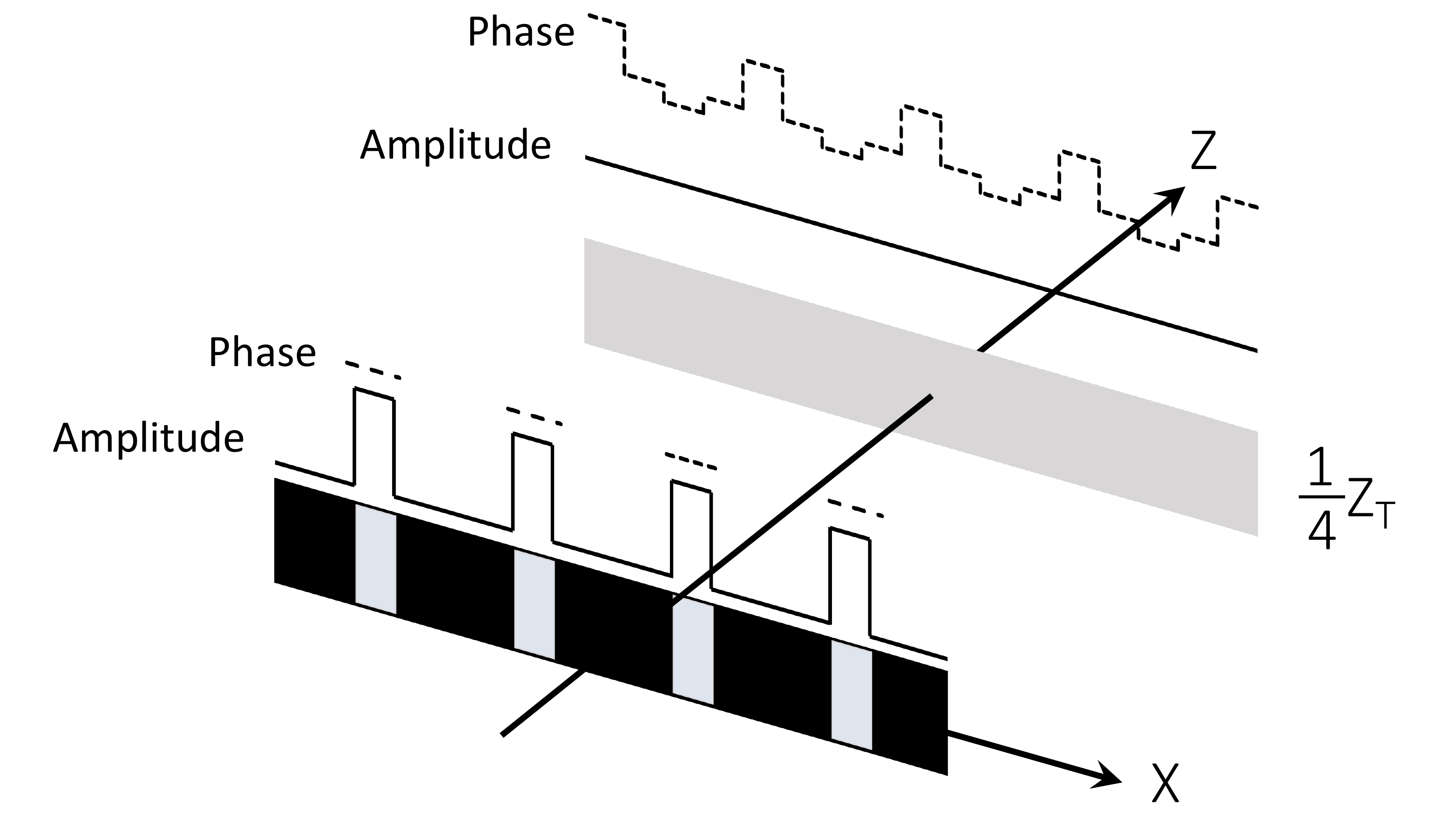}
\caption{\label{finvTAI} Scheme of the inverse Talbot array illuminator for $p/q =1/4$.}
\end{figure}

The TAI phases can be determined when the propagation is considered in the reversed way \cite{szwaykowski1993talbot},\cite{arrizon1994multilevel}, as shown in Fig. \ref{finvTAI}. Starting from a binary amplitude grating of opening ratio $1/q$, which is illuminated by a collimated and spatially coherent wavefront, one searches for distributions of uniform light intensity at the fractional Talbot planes after the amplitude grating. At these planes the wavefront is composed of a set of phase levels with a periodicity equal to the original period $L$ of the amplitude grating. In this inverse TAI scheme, the design problem is simply to determine these phase levels at the fractional Talbot planes. Note that in the spatial formalism reversing the propagation direction $z$ is equivalent to a complex conjugation, since the one-dimensional paraxial wave equation is:
\begin{equation}
2ik_0\frac{\partial A}{\partial z} = -\frac{\partial^2 A}{\partial x^2}, 
\end{equation}
where $A(x,z)$ is the complex amplitude of the paraxial wave and $k_0$ the wavenumber. Therefore, the TAI phases are the complex conjugated to those acquired in propagation. 

Alternatively, one can reproduce in a multilevel phase grating the phases acquired by fractional $p/q$ Talbot propagation, without complex conjugation, and operate the phase grating at the complementary $(q-p)/q$ fractional Talbot distance \cite{arrizon1994multilevel}. The combined effect is the propagation over a $q/q =1$ Talbot distance, and thus the original amplitude grating is reproduced with a half-period shift. In our presentation, however, we use the former point of view as it does not require the change of the Talbot order. 

Within the time-domain formalism the general solution can be found as follows. As in the inverse TAI problem, let us consider the temporal analogue to a binary amplitude grating with opening ratio $1/q$, which is a pulse of rectangular form and width $T/q$.  Using (\ref{tproc}), we observe that the output unit cell at a fractional Talbot plane of order $p/q$ is composed of $q$ replicas of the basic rectangular pulse, each with a quadratic phase factor proportional to the ratio $s/q$. According to the previous development, the TAI phases should be the complex conjugates. Explicitly,  
\begin{align}\label{TAIp}
\Phi^{(s/q)}_n= \exp\left(-j\sigma_\phi \xi_n\right)=\exp\left[-j\sigma_\phi\left(\xi_0+\pi\frac{s}{q} n^2\right)\right].
\end{align}
To check this result, let us consider the direct TAI geometry. Cw laser light of unit amplitude, analogous to the plane wave that illuminates the grating in Fig. \ref{fTAI}, is phase-modulated according to the levels in (\ref{TAIp}), each in a time interval of duration $T/q$. The input unit cell can be presented as: 
\begin{align}\label{TAIr}
w\left(t\right)=\sum_{n=0}^{q-1} \exp\left(-j\sigma_\phi\xi_n\right) 
{\rm rect}\left(\frac{tq}{T}-n\right),
\end{align}
where the rectangle function is defined as ${\rm rect}(x)=1$ for $|x|<1/2$ and zero otherwise. To compute the field after fractional Talbot we use (\ref{tproc}) and (\ref{TAIr}) in (\ref{tte}), obtaining:
\begin{align}\label{TAIf}
&E'(t)=\frac{1}{\sqrt{q}}\sum_{m=-\infty}^{+\infty}
\sum_{n=0}^{q-1}e^{j\sigma_\phi(\xi_m-\xi_n)}
{\rm rect}\left(\frac{tq}{T}-n-m\right).
\end{align}
We have omitted the $\delta(t-e_{pq}T)$ term for simplicity, since it is not relevant in the present computation. With the change $n+m=k$ this reads:
\begin{align}\label{TAIf2}
&E'(t)=\frac{1}{\sqrt{q}}\sum_{k=-\infty}^{+\infty}
\sum_{n=0}^{q-1}e^{j\sigma_\phi(\xi_{k-n}-\xi_n)}
{\rm rect}\left(\frac{tq}{T}-k\right).
\end{align}
The sum in $n$ is in the form of a convolution, but can be written as the autocorrelation $R_\xi(-k)=q\delta_{k,0\ (\mathrm{mod}\ q)}$ using the symmetry property (\ref{properties}). The autocorrelation is nonzero only for $k=0$ (moq $q$), so setting $k=nq$ for $n$ integer we finally get:
\begin{align}\label{f}
E'(t)
=\sqrt{q}\sum_{n=-\infty}^{+\infty}
{\rm rect}\left(\frac{tq}{T}-nq\right).
\end{align}
The output envelope is the expected rectangle pulse train with period $T$ and pulse width $T/q$. 

\begin{figure*}[ht]
\includegraphics[scale=0.4]{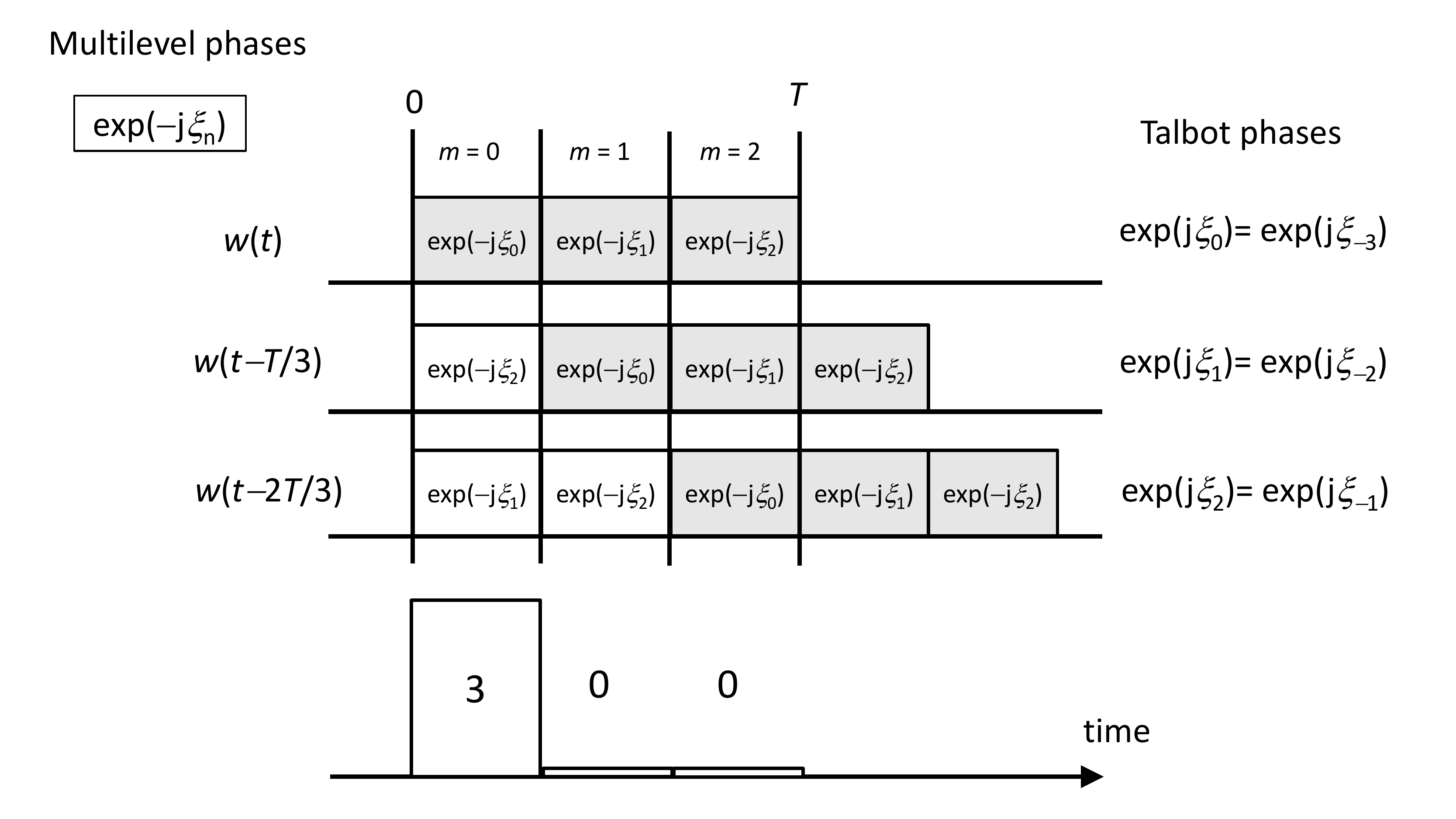}
\caption{\label{bins} Scheme of the multiple interference in time bins in a Talbot array illuminator with $p/q =1/3$ and $\sigma_\phi = +1$. The unit cell is contained in interval $[0,T)$ and divided in three time bins, shown with a gray background in the top row. Each bin contains the corresponding multilevel phase. The shifting (left) and phase weighting (right) of bins induced by Talbot effect is shown in successive rows, using the cyclic structure of bins inherited by the periodicity of the unit cell. The result of the bin-wise coherent sum is represented in the bottom.}
\end{figure*}

This result can be described as follows. The multilevel phase modulation (\ref{TAIr}) defines a series of $q$ time bins of duration $T/q$ within the fundamental period $T$, explicitly shown by the succession of rect functions. The shifting and weighting of unit cells produced by Talbot effect, together with the periodicity of the the train, translates into a shifting and weighting of bins, which leads to the convolution expressed in Eq. (\ref{TAIf}). This process has been exemplified in Fig. \ref{bins}. If the multilevel phases are chosen as the complex conjugated to the Talbot propagation phases, the bin-wise multiple interference becomes the perfect periodic autocorrelation (\ref{auto}). The multiple interference is therefore constructive for only only bin per period, and is destructive for the rest. In other words, the array illuminator relies on the autocorrelation properties of the quadratic Gauss sum arising in Talbot effect. 

The TAI operation principle can also be analyzed in the spectral domain. The Fourier transform of the unit cell (\ref{TAIr}) is:
\begin{align}\label{TAIr2}
W\left(\nu\right)=\frac{T}{q}\mathrm{sinc}\left(\frac{\nu T}{q}\right) \sum_{n=0}^{q-1} e^{-j\sigma_\phi\xi_n}
e^{-j2\pi\nu nT/q}.
\end{align}
Since the unit cell is repeated with period $T$, the spectrum of the envelope $E(t)$ is composed of spectral lines $\nu=m/T$, with $m$ integer:
\begin{align}
E(\nu)=&\frac{1}{q}
\mathrm{sinc}\left(\frac{\nu T}{q}\right)\sum_{m=-\infty}^{+\infty}
\delta\left(\nu-\frac{m}{T}\right)\times \nonumber\\
&\sum_{n=0}^{q-1}e^{-j\sigma_\phi\xi_n}
e^{-j2\pi {m n}/{q}},
\end{align}
where sinc($x$)=sin($\pi x$)/($\pi x$). The sum over $n$ becomes the DFT (\ref{gausspair}), and the result can be presented as:
\begin{align}
E(\nu)
=&\frac{1}{\sqrt{q}}\exp\left[j\pi \sigma_\phi\frac{p}{q}(\nu T)^2\right]\times\nonumber\\
&\mathrm{sinc}\left(\frac{\nu T}{q}\right)
\sum_{m=-\infty}^{+\infty}
\delta\left(\nu-\frac{m}{T}\right)(-1)^{pqm}.
\end{align}
This is the spectrum of a train of squared pulses of width $T/q$, amplitude $\sqrt{q}$ and period $T$, shifted by half a period when $pq$ is odd, and dispersed by the quadratic phase factor in the first line. The dispersive line after the phase modulator simply compensates for this phase factor, and renders the pulse train chirp-free and thus of square temporal profile. 

The peak amplitude in (\ref{f}) is $\sqrt{q}$, which amounts to a gain factor of $q$ in peak power with respect to the power of the cw laser light. Notice also that (\ref{TAIr}) can be interpreted as the multilevel phase modulation of a perfect rectangular pulse. In general, had we phase modulated an arbitrary pulse contained in an interval of duration $T/q$ and repeated also in $T/q$ shifts, we would have obtained at the output the same pulse with periodicity $T$ and a gain factor of $q$ in power. This is the principle of the noiseless pulse amplification by coherent addition demonstrated in \cite{reza2014noiseless}.

It is illustrative to analyze some particular cases derived from (\ref{TAIp}). For instance, it contains the six binary TAIs described in \cite{arrizon1993talbot}, associated to the fractional Talbot planes $p/q=1/2$ ($s=1$), $p/q=1/3$ ($s=4$), $p/q=2/3$ ($s=2$), and the corresponding shifted integers of the form $(p+q)/q$. Moreover, for $p=1$, and using the first result of Proposition 2, phases (\ref{TAIp}) take the form:
\begin{equation}\label{TAIp1}
\Phi^{(1/q)}_n=(-1)^{ne_q} \exp(j\sigma_\phi\pi n^2/q),
\end{equation}
which coincide, up to a global phase, with the series of solutions originally presented in \cite{leger1990efficient}. The equivalence is immediate for $q$ even, and for $q$ odd, and using their notation, it follows from the substitution $n=I+(q/2)$ with $I$ half-integer ranging from $-N/2$ to $(N/2)-1$. 

We also mention that the set of TAI phases (\ref{TAIp}) is not the same set of phases used to induce the spectral Talbot effect \cite{azana2005spectral} by use of multilevel phase modulation \cite{caraquitena2011spectral}. The spectral Talbot effect consists in the generation of self-imaging phenomena in the spectrum of a frequency comb, {\it i. e.}, in the broadband spectrum of a train of ultrashort pulses of period $T$, where the set of spectral lines separated by a frequency difference of $1/T$ plays the role of a diffraction grating in the usual configuration of the Talbot effect. In order to induce the spectral Talbot effect it is thus necessary to generate a quadratic phase transformation in the reciprocal domain, in this case in time. Such a transformation is therefore a quadratic phase modulation of the form $\exp(j\alpha t^2/2)$ for certain values of the chirp parameter $\alpha$. For ultrashort pulses, which are essentially located at definite temporal instants $t_n =nT$, the spectral Talbot effect is generated simply by setting $\alpha =\pm 2\pi\frac{p}{q}T^{-2}$. Therefore, the effect can be induced by the set of phases \cite{caraquitena2011spectral}: 
\begin{align}\label{STp}
\Phi^{(p/q)}_n=\exp\left(\pm j\pi \frac{p}{q} n^2\right).
\end{align}
This sequence of phases has, in general, a period of $2q$, contrary to (\ref{TAIp}) where the period is always $q$. This is because in (\ref{STp}) ratio $p/q$, which is the order of the (spectral) Talbot effect, can contain two odd numbers. By contrast, ratio $s/q$ in (\ref{TAIp}) is an irreducible fraction of integers with opposite parity, as shown by the third property of Theorem 1. Notice, however, that the two sets of phases coincide at the self-dual integers $s=p=q\pm 1$.

In the rest of the section we establish the connection with a known characterization of the TAI phases up to an irrelevant global phase and an arbitrary shift of the phase grating by integer multiples of $T/q$ \cite{zhou1999analytic}, which provide an alternative form of Chu's construction \citep{chu1972polyphase}. This result is thus sufficient when the illumination shows this symmetry, as is the case of the TAI problem, but not in more complex systems where this symmetry is broken, such as in Talbot lines with structured illumination.

The TAI phases are written in \cite{zhou1999analytic} in terms of the modular inverse of $p$ (mod $q$), denoted as $r$:
\begin{equation}
r=\left[\frac{1}{p}\right]_q. 
\end{equation}
We first analyze the relationship between this $r$ and our $s$. We notice that, although in some cases they coincide, these integers may differ. For instance, for $p=5$ and $q=8$ we have $r=[1/5]_8 = 5$ but $s=[1/5]_{16} = 13$. In general, integers $r$ and $s$ are either equal (mod $2q$) or differ by $q$, since the condition $pr = 1\ ({\rm mod }\ q)$ implies either $s = r\ ({\rm mod } \ 2q)$ or $s = r + q\ ({\rm mod } \ 2q)$. For $q$ even this means that, since $s$ must be odd, $r$ is also odd, independently of the concrete form of the relationship, $s=r$ or $s=r+q$. For $q$ odd integer $s$ must be even, and so $s=r+q$ is allowed only when $r$ is odd, {\it i.e.}, $s = r +qe_r \ ({\rm mod }\ 2q)$, since this assures that $s$, as a function of $r$, is always even. We are now in position to express the connection:\\

{\bf Proposition 5.} The TAI phases can be written as:
\begin{align}\label{tek7}
\exp\left(j\pi\frac{s}{q}n^2\right)&\sim \exp\left(j\pi\frac{r}{q}m^2\right)\ \  (q \ {\rm even})\nonumber\\
\exp\left(j\pi\frac{s}{q}n^2\right)&\sim \exp\left(j\pi\frac{r}{q}m(m-1)\right)\ \ (q\  {\rm odd}),
\end{align}\\
where $\sim$ stands for equality up to a global phase and a possible shift of index $n$.\\

{\bf Proof.} We must analyze case by case. For $q$ even, the equality is obvious if $s=r$, and for $s=r+q$ we have:
\begin{align}\label{tek6}
&\exp\left(j\pi\frac{s}{q}n^2\right)=
 \exp\left(j\pi\frac{r}{q}n^2\right)(-1)^n	 \nonumber \\
=&\exp\left(j\pi\frac{r}{q}n^2\right)(-1)^{nr}= 
\exp\left(j\pi\frac{r}{q}n(n-q)\right) \nonumber\\
=&\exp\left(j\pi\frac{r}{q} \left(n-\frac{q}{2}
\right)^2 \right) \exp\left( -j\pi\frac{qr}{4}\right),
\end{align}
where in the second step we have used that $r$ is odd. The equivalence up to a global phase follows after the shift $m=n-({q}/{2})$.

For $q$ odd we use the explicit formula derived above, $s=r+qe_r$ and compute for both cases. For $r$ odd or $s=r+q$, and using the same algebra as in (\ref{tek6}), we have:
\begin{align}\label{tek8}
&\exp\left(j\pi\frac{s}{q}n^2\right) 
= 
\exp\left(j\pi\frac{r}{q}n(n-q)\right)= \nonumber\\
&\exp\left( j\pi\frac{r}{q}
m(m-1)\right) \exp\left( -j\pi\frac{q^2-1}{4}\right)
\end{align}
after the shift $m=n-[{(q-1)}/{2}]$, showing the equivalence with (\ref{tek7}). Finally, for $r$ even or $s=r$ we have: 
\begin{align}\label{tek10}
\exp \left(j\pi\frac{s}{q}n^2\right) = \exp\left(j\pi\frac{r}{q}n^2\right)
 =\exp\left(j\pi\frac{r}{q}n(n-q)\right),
\end{align}
because the additional term $\exp(-j\pi rn) = (-1)^{rn}$ is unity since $r$ is even. In this form, (\ref{tek10}) is equivalent to (\ref{tek8}) and leads again to (\ref{tek7}). The proof is complete. 

\section{Conclusions}

In this paper we have presented a novel and simple computation of the quadratic Gauss sums that define the Talbot effect. It is based on the properties of integer $s$, a parity-dependent modular inverse constructed from the two coprime numbers $p$ and $q$ that define a general fractional Talbot plane. The computation can be compactly described in two cases depending on the parity of $q$, and shows that the DFT of a quadratic phase sequence is itself a quadratic phase sequence. In addition, both members of the DFT pair have perfect periodic autocorrelation. These results were exemplified by the study of Talbot array illuminators, where the perfect autocorrelation arises as a multiple interference process in bins within the fundamental period of the grating. Also, connection with previous characterizations of TAIs have been provided. The results derived here represent a simple means for the design and analysis of systems based on fractional Talbot effect.

\begin{acknowledgments}
This paper was written during a stay at INRS-EMT, Montreal, Canada, funded by Generalitat Valenciana, Spain through a BEST/2016/281 grant, and evolved from conversations with Reza Maram, Luis Romero Cort\'es and Jos\'e Aza{\~n}a, to whom I am indebted. 

This paper is dedicated to the memory of Profs. Mar\'ia Victoria P\'erez, Carlos G\'omez-Reino and Felipe Mateos, with whom I learnt the beauty of Talbot effect.
\end{acknowledgments}

\appendix
\section{\label{appnt}Some results from number theory}

A basic result in number theory \cite{Hecke}, \cite{Lang}, \cite{IrelandRosen}, sometimes referred to as B\'ezout's lemma, states that given two integers, $a$ and $b$, with greatest common divisor $d$, the modular equation $ax+by=n$, is solvable for $x$ and $y$ if and only if integer $n$ is multiple of $d$. In particular, if $a$ and $b$ are coprime, the result guarantees that there exist solutions to the equation:
\begin{equation}\label{bezout}
ax+by=1.
\end{equation} 
The pairs of solutions $(x_0,y_0)$ of (\ref{bezout}) are not unique, and in fact determined mod $b$ and $a$, respectively, since $(x_0+mb,y_0-ma)$ for any integer $m$ is also a pair of solutions. It can be shown that this series exhaust the possible pairs of solutions to (\ref{bezout}). Let us define $x_0$ as the lowest positive integer that belongs to a pair of solutions to (\ref{bezout}). $x_0$ is non-zero and bounded, $x_0 < b$. The modular multiplicative inverse, or simply the modular inverse of $a$ (mod $b$), denoted as $[1/a]_b$, is this unique integer $x_0$ that verifies: 
\begin{equation}\label{modinv}
a\left[\frac{1}{a}\right]_b = 1\ ({\rm mod}\ b)
\end{equation} 
and lies in the range $1 \leq [1/a]_b \leq b-1$. The same construction can be applied to the modular inverse $[1/b]_a$, which in this case lies in the range $1\leq [1/b]_a \leq a-1$. 

B\'ezout's lemma also implies that, given $a$ and $b$ coprime, the equation $ax+by=n$ admits solutions for any integer $n$. In particular, this means that the modular equation $ax_n=n$ (mod $b$) is solvable for any $n$ \cite[p. 31]{IrelandRosen}. A complete set of solutions is given by $x_n=n[1/a]_b$ (mod $b$), including the case $n=0$. If we set $n$ in the range of residues mod $b$, $0\leq n\leq b-1$ and reduce the values of $x_n$ to their residues mod $b$, we have thus defined a bijective map $n\leftrightharpoons x_n$ within the residues mod $b$. This is the map that has been used to perform changes of variables of the type $x\rightarrow y=ax$ (mod $b$) for $a$, $b$ coprime in invariant expressions mod $b$ in several parts of the paper.

The Jacobi symbol $\left(\frac{a}{b}\right)$ of an arbitrary integer $a$ and an odd and positive integer $b$, which are mutually prime, is defined as the product of the Legendre symbols of the prime factors of $b$. To be specific, if the decomposition of $b$ in prime factors is $b=p_1^{\alpha_1}\cdots p_N^{\alpha_N}$, then:
\begin{align}\label{Jacobi}
\left(\frac{a}{b}\right) =\left(\frac{a}{p_1}\right)^{\alpha_1}\cdots\left(\frac{a}{p_N}\right)^{\alpha_N}
\end{align}
where for prime $p_j$, the Legendre symbol is: 
\begin{align}
\left(\frac{a}{p_j}\right) =\begin{cases}
+1&\textrm{if there exists an integer }\\
& x\textrm{ such that } x^2=a\ (\textrm{mod } p_j)\\
-1&\textrm{otherwise.}\\
\end{cases} 
\end{align}
Acoording to this definition, the Legendre symbol $(a/p_j)$ equals $+1$ if and only if $a$ is a quadratic residue mod $p_j$, otherwise it is $-1$. In particular, $(1/p_j) = +1$ for any $p_j$, and $(a/p_j) = (a'/p_j)$ if $a=a'$ (mod $p_j$). Properties and values of the Legendre symbols can be consulted in \cite{Hecke}, \cite{Lang}, \cite{IrelandRosen} and in the second appendix of \cite{HB80}. The Jacobi symbol (\ref{Jacobi}) verifies the same properties as the Legendre symbols, and also verifies that its value is $+1$ when $a$ is a quadratic residue mod $b$. The converse, however, is not necessarily true, since for $a$ being a quadratic residue of $b$ it is necessary that $a$ is quadratic residue of all of its prime factors $p_j$, and so all factors in (\ref{Jacobi}) must be $+1$.

\section{{\label{appeq}}Equivalence with alternative expressions of integer $s$} 

The computation of $\exp(j\xi_n)$ in \citep{HB80}, corrected in \cite{MO03} and as presented in \cite{LHA15}, is similar to (\ref{gaussphase}) by with a different expression for $s$. We show here the equivalence of the integer $s$ in (\ref{s-qpar}) and (\ref{s-qimpar}) with the residues mod $2q$ of the three types of integers defined in \cite{MO03}, \cite{LHA15}, here denoted as $s'$, by showing that they verify the conditions stated in Theorem 1. In the first case we also show the connection by explicit manipulations on $s\prime$. \\

\textbf{(a) $p$ odd and $q$ odd.} The integer is given by: 
\begin{equation}\label{tek4}
s' =8p\left[\frac{1}{2}\right]_q\left[\frac{1}{2p}\right]_q^2 \ \ ({\rm mod} \ 2q).
\end{equation}
Using that $\left[1/2\right]_q=(q+1)/2$ \cite{HB80} \cite{Lang}, we can simplify this expression to:
\begin{equation}\label{tek1}
s'=4p\left[\frac{1}{2p}\right]_q^2 \ \ ({\rm mod} \ 2q).
\end{equation}
Incidentally, we mention that the value $s'$ in the original computation \citep{HB80} was:
\begin{equation}
s'_{HB}=4p\left[\frac{1}{4p}\right]_q^2 \ \ ({\rm mod} \ 2q).
\end{equation}
Returning to (\ref{tek1}), we recall the definition of the modular inverse:
\begin{equation}\label{tek3}
2p\left[\frac{1}{2p}\right]_q =1 \ \ ({\rm mod} \ q), 
\end{equation}
we can be further simplify (\ref{tek1}) to: 
\begin{align}
s' &=2\left[\frac{1}{2p}\right]_q \times[1 \, ({\rm mod} \,q)]  \ \ ({\rm mod} \ 2q)\nonumber\\
&=2\left[\frac{1}{2p}\right]_q \ \ ({\rm mod} \ 2q),
\end{align}
which coincides with (\ref{s-qimpar}). 

Alternatively, we can show the equivalence by a direct check of the hypotheses of Theorem 1. First we rephrase (\ref{tek3}) as:
\begin{equation}
2p\left[\frac{1}{2p}\right]_q = 1 + aq
\end{equation}
for certain integer $a$, which should be odd since $q$ is odd and the left hand side of the equation is even. Then the product $s' p$ is:
\begin{equation}
s' p=\left(2p\left[\frac{1}{2p}\right]_q\right)^2 
= (1+aq)^2 = 1 + q \ \ ({\rm mod} \ 2q),
\end{equation} 
which is the defining equation (\ref{lmdc}) in Theorem 1. This second form of the equivalence is completed by observing in (\ref{tek4}) that the parity of $s'$ is even, and is thus opposite to that of $q$.\\

\textbf{(b) $p$ odd and $q$ even.} The integer is defined as: 
\begin{equation}\label{tek5}
s' =p\left[\frac{1}{p}\right]_q^2 \ \ ({\rm mod} \ 2q).
\end{equation}
Using the definition of modular inverse,
\begin{equation}\label{tek2}
p\left[\frac{1}{p}\right]_q = 1 \ ({\rm mod} \ q)= 1 + bq
\end{equation}
for certain integer $b$, we have that the product $s' p$ is: 
\begin{equation}
s' p=(1 + bq)^2= 1 + 2bq +(bq)^2 = 1\ \ ({\rm mod}\ 2q),
\end{equation}
since $q$ is even. This is the defining equation (\ref{lmdc}) for $q$ even. We finally notice from (\ref{tek2}) that $\left[1/p\right]_q$ must be odd, so $s'$ is odd and thus of opposite parity to $q$.\\

\textbf{(c) $p$ even and $q$ odd.} The integer is:
\begin{equation}
s' =p\left[\frac{1}{p}\right]_q^2 \ \ ({\rm mod} \ 2q).
\end{equation}
The definition of modular inverse (\ref{tek2}) now implies that integer $b$ is odd, since $q$ is odd and $p$ is even. Then, the product $s' p$ is:
\begin{equation}
s' p=(1 + bq)^2= 1 + q \ \ ({\rm mod}\ 2q),
\end{equation}
which is again the defining equation (\ref{lmdc}) for $q$ odd. The proof is completed by noticing that $s'$ is now even since $p$ is even.

\section{\label{appphase}Computation of phase $exp(j\xi_0)$}

We compute the Gauss sum (\ref{gaussp3}), which leads to the result (\ref{gaussphase}). As in \cite{HB80}, the computation is reduced to the application of the standard form of the Gauss sum found in texts of number theory \cite[p. 86]{Lang}. This standard form applies to sums of quadratic phases of the type $\exp(j2\pi an^2/b)$ where $a$ is an arbitrary integer and $b$ is odd. In the case of $q$ odd, we are in these conditions since $s$ is even. Then, 
\begin{align}
e^{j\xi_0} &= \left(\frac{s/2}{q}\right) e^{j\pi(q-1)^2/8}\nonumber\\
&=\left(\frac{s}{q}\right) \left(\frac{2}{q}\right) e^{j\pi(q-1)^2/8}=
\left(\frac{s}{q}\right) e^{j\pi(q-1)/4},
\end{align}
where in the second step we have used the numerator product rule of Jacobi symbols and the value $(2/q) = \exp[-j\pi(q^2-1)/8]$. This gives the first formula in (\ref{gaussphase}). Now, we use (\ref{lmdc}) and again the numerator product rule to reach an expression involving $p$ and $q$: 
\begin{align}
\left(\frac{s}{q}\right)\left(\frac{p}{q}\right) =\left(\frac{sp}{q}\right)=\left(\frac{1+qe_q}{q}\right)=\left(\frac{1}{q}\right) = 1,
\end{align}
and so $(s/q)=(p/q)$ and the first line of (\ref{gaussphase}) is proved.

For $q$ even, $s$ is odd and we have to invert the fraction in the exponential of (\ref{gaussp3}). Using the reciprocity of the Gauss sums \cite{HB80}, \citep{MO03}, \cite{Hecke} we are led to:
\begin{equation}\label{duality}
e^{j\xi_0} =\frac{1}{\sqrt{q}} \sum_{m=0}^{q-1} \, e^{-j\pi \frac{s}{q}m^2}=
\frac{e^{ -j\pi /4}}{\sqrt{s}} \sum_{m=0}^{q-1} \, e^{ j\pi \frac{q}{s}m^2}.
\end{equation}
The use of the standard form gives the first expression of (\ref{gaussphase}) for $q$ even: 
\begin{align}\label{sum2}
e^{j\xi_0} =\left( \frac{q}{s}\right) e^{-j\pi s/4}.
\end{align}
Since the original Gauss sum in (\ref{duality}) is invariant under shifts $s\rightarrow s+2q$, so it is the right hand side of (\ref{sum2}), and therefore:
\begin{align}
 \left( \frac{q}{s+2qn}\right)=\left( \frac{q}{s}\right) (-1)^{qn/2}.
\end{align}
for any $n\geq 0$. In particular, we can evaluate (\ref{sum2}) using any other representation of $s$ equivalent mod $2q$. With the representation $s=p[1/p]_q^2$ from Appendix \ref{appeq} we can straightforwardly reduce this result to an expression involving $p$ and $q$. First, 
\begin{align}\label{sum3}
\left(\frac{q}{s}\right)=\left(\frac{q}{p}\right)\left(\frac{q}{[1/p]_q}\right)^2 = \left(\frac{q}{p}\right), 
\end{align}
where we have used the denominator product rule of Jacobi symbols and the fact that they only take values $\pm 1$. And second, we observe that since $s=p[1/p]_q^2$ is odd, $[1/p]_q$ must also be odd. Therefore, setting $s = p (2t+1)^2$ for certain integer $t$ we get:
\begin{align}
e^{-j\pi s/4}= e^{-j\pi p(2t+1)^2/4}= e^{-j\pi p/4},
\end{align}
which together with (\ref{sum3}) gives the second expression in (\ref{gaussphase}) for $q$ even.

\bibliography{Gauss}

\begin{thebibliography}{28}%
\makeatletter
\providecommand \@ifxundefined [1]{%
 \@ifx{#1\undefined}
}%
\providecommand \@ifnum [1]{%
 \ifnum #1\expandafter \@firstoftwo
 \else \expandafter \@secondoftwo
 \fi
}%
\providecommand \@ifx [1]{%
 \ifx #1\expandafter \@firstoftwo
 \else \expandafter \@secondoftwo
 \fi
}%
\providecommand \natexlab [1]{#1}%
\providecommand \enquote  [1]{``#1''}%
\providecommand \bibnamefont  [1]{#1}%
\providecommand \bibfnamefont [1]{#1}%
\providecommand \citenamefont [1]{#1}%
\providecommand \href@noop [0]{\@secondoftwo}%
\providecommand \href [0]{\begingroup \@sanitize@url \@href}%
\providecommand \@href[1]{\@@startlink{#1}\@@href}%
\providecommand \@@href[1]{\endgroup#1\@@endlink}%
\providecommand \@sanitize@url [0]{\catcode `\\12\catcode `\$12\catcode
  `\&12\catcode `\#12\catcode `\^12\catcode `\_12\catcode `\%12\relax}%
\providecommand \@@startlink[1]{}%
\providecommand \@@endlink[0]{}%
\providecommand \url  [0]{\begingroup\@sanitize@url \@url }%
\providecommand \@url [1]{\endgroup\@href {#1}{\urlprefix }}%
\providecommand \urlprefix  [0]{URL }%
\providecommand \Eprint [0]{\href }%
\providecommand \doibase [0]{http://dx.doi.org/}%
\providecommand \selectlanguage [0]{\@gobble}%
\providecommand \bibinfo  [0]{\@secondoftwo}%
\providecommand \bibfield  [0]{\@secondoftwo}%
\providecommand \translation [1]{[#1]}%
\providecommand \BibitemOpen [0]{}%
\providecommand \bibitemStop [0]{}%
\providecommand \bibitemNoStop [0]{.\EOS\space}%
\providecommand \EOS [0]{\spacefactor3000\relax}%
\providecommand \BibitemShut  [1]{\csname bibitem#1\endcsname}%
\let\auto@bib@innerbib\@empty
\bibitem [{\citenamefont {Talbot}(1836)}]{talbot1836}%
  \BibitemOpen
  \bibfield  {author} {\bibinfo {author} {\bibfnamefont {H.~F.}\ \bibnamefont
  {Talbot}},\ }\href@noop {} {\bibfield  {journal} {\bibinfo  {journal}
  {Philosophical Magazine}\ }\textbf {\bibinfo {volume} {9}},\ \bibinfo {pages}
  {401} (\bibinfo {year} {1836})}\BibitemShut {NoStop}%
\bibitem [{\citenamefont {Rayleigh}(1881)}]{rayleigh1881}%
  \BibitemOpen
  \bibfield  {author} {\bibinfo {author} {\bibfnamefont {L.}~\bibnamefont
  {Rayleigh}},\ }\href@noop {} {\bibfield  {journal} {\bibinfo  {journal}
  {Philosophical Magazine}\ }\textbf {\bibinfo {volume} {11}},\ \bibinfo
  {pages} {196} (\bibinfo {year} {1881})}\BibitemShut {NoStop}%
\bibitem [{\citenamefont {Winthrop}\ and\ \citenamefont
  {Worthington}(1965)}]{winthrop1965}%
  \BibitemOpen
  \bibfield  {author} {\bibinfo {author} {\bibfnamefont {J.~T.}\ \bibnamefont
  {Winthrop}}\ and\ \bibinfo {author} {\bibfnamefont {C.~R.}\ \bibnamefont
  {Worthington}},\ }\href@noop {} {\bibfield  {journal} {\bibinfo  {journal}
  {Journal of the Optical Society of America}\ }\textbf {\bibinfo {volume}
  {55}},\ \bibinfo {pages} {373} (\bibinfo {year} {1965})}\BibitemShut
  {NoStop}%
\bibitem [{\citenamefont {Aza{\~n}a}\ and\ \citenamefont
  {de~Chatellus}(2014)}]{azana2014angular}%
  \BibitemOpen
  \bibfield  {author} {\bibinfo {author} {\bibfnamefont {J.}~\bibnamefont
  {Aza{\~n}a}}\ and\ \bibinfo {author} {\bibfnamefont {H.~G.}\ \bibnamefont
  {de~Chatellus}},\ }\href@noop {} {\bibfield  {journal} {\bibinfo  {journal}
  {Physical Review Letters}\ }\textbf {\bibinfo {volume} {112}},\ \bibinfo
  {pages} {213902} (\bibinfo {year} {2014})}\BibitemShut {NoStop}%
\bibitem [{\citenamefont {Akhmanov}\ \emph {et~al.}(1969)\citenamefont
  {Akhmanov}, \citenamefont {Sukhorukov},\ and\ \citenamefont
  {Chirkin}}]{akhmanov1969nonstationary}%
  \BibitemOpen
  \bibfield  {author} {\bibinfo {author} {\bibfnamefont {S.}~\bibnamefont
  {Akhmanov}}, \bibinfo {author} {\bibfnamefont {A.}~\bibnamefont
  {Sukhorukov}}, \ and\ \bibinfo {author} {\bibfnamefont {A.}~\bibnamefont
  {Chirkin}},\ }\href@noop {} {\bibfield  {journal} {\bibinfo  {journal}
  {Soviet Journal of Experimental and Theoretical Physics}\ }\textbf {\bibinfo
  {volume} {28}},\ \bibinfo {pages} {748} (\bibinfo {year} {1969})}\BibitemShut
  {NoStop}%
\bibitem [{\citenamefont {Kolner}(1994)}]{kolner1994space}%
  \BibitemOpen
  \bibfield  {author} {\bibinfo {author} {\bibfnamefont {B.~H.}\ \bibnamefont
  {Kolner}},\ }\href@noop {} {\bibfield  {journal} {\bibinfo  {journal} {IEEE
  Journal of Quantum Electronics}\ }\textbf {\bibinfo {volume} {30}},\ \bibinfo
  {pages} {1951} (\bibinfo {year} {1994})}\BibitemShut {NoStop}%
\bibitem [{\citenamefont {Aza{\~n}a}\ and\ \citenamefont
  {Muriel}(2001)}]{azana2001temporal}%
  \BibitemOpen
  \bibfield  {author} {\bibinfo {author} {\bibfnamefont {J.}~\bibnamefont
  {Aza{\~n}a}}\ and\ \bibinfo {author} {\bibfnamefont {M.~A.}\ \bibnamefont
  {Muriel}},\ }\href@noop {} {\bibfield  {journal} {\bibinfo  {journal} {IEEE
  Journal of Selected Topics in Quantum Electronics}\ }\textbf {\bibinfo
  {volume} {7}},\ \bibinfo {pages} {728} (\bibinfo {year} {2001})}\BibitemShut
  {NoStop}%
\bibitem [{\citenamefont {Aza{\~n}a}(2005)}]{azana2005spectral}%
  \BibitemOpen
  \bibfield  {author} {\bibinfo {author} {\bibfnamefont {J.}~\bibnamefont
  {Aza{\~n}a}},\ }\href@noop {} {\bibfield  {journal} {\bibinfo  {journal}
  {Optics Letters}\ }\textbf {\bibinfo {volume} {30}},\ \bibinfo {pages} {227}
  (\bibinfo {year} {2005})}\BibitemShut {NoStop}%
\bibitem [{\citenamefont {Caraquitena}\ \emph {et~al.}(2011)\citenamefont
  {Caraquitena}, \citenamefont {Beltr{\'a}n}, \citenamefont {Llorente},
  \citenamefont {Mart{\'\i}},\ and\ \citenamefont
  {Muriel}}]{caraquitena2011spectral}%
  \BibitemOpen
  \bibfield  {author} {\bibinfo {author} {\bibfnamefont {J.}~\bibnamefont
  {Caraquitena}}, \bibinfo {author} {\bibfnamefont {M.}~\bibnamefont
  {Beltr{\'a}n}}, \bibinfo {author} {\bibfnamefont {R.}~\bibnamefont
  {Llorente}}, \bibinfo {author} {\bibfnamefont {J.}~\bibnamefont
  {Mart{\'\i}}}, \ and\ \bibinfo {author} {\bibfnamefont {M.~A.}\ \bibnamefont
  {Muriel}},\ }\href@noop {} {\bibfield  {journal} {\bibinfo  {journal} {Optics
  Letters}\ }\textbf {\bibinfo {volume} {36}},\ \bibinfo {pages} {858}
  (\bibinfo {year} {2011})}\BibitemShut {NoStop}%
\bibitem [{\citenamefont {Patorski}(1989)}]{patorski89}%
  \BibitemOpen
  \bibfield  {author} {\bibinfo {author} {\bibfnamefont {K.}~\bibnamefont
  {Patorski}},\ }\enquote {\bibinfo {title} {The self-imaging phenomenon and
  its applications},}\ in\ \href@noop {} {\emph {\bibinfo {booktitle} {Progress
  in Optics}}},\ Vol.~\bibinfo {volume} {27},\ \bibinfo {editor} {edited by\
  \bibinfo {editor} {\bibfnamefont {E.}~\bibnamefont {Wolf}}}\ (\bibinfo
  {publisher} {Elsevier},\ \bibinfo {year} {1989})\ pp.\ \bibinfo {pages} {1 --
  108}\BibitemShut {NoStop}%
\bibitem [{\citenamefont {Wen}\ \emph {et~al.}(2013)\citenamefont {Wen},
  \citenamefont {Zhang},\ and\ \citenamefont {Xiao}}]{wen13talbot}%
  \BibitemOpen
  \bibfield  {author} {\bibinfo {author} {\bibfnamefont {J.}~\bibnamefont
  {Wen}}, \bibinfo {author} {\bibfnamefont {Y.}~\bibnamefont {Zhang}}, \ and\
  \bibinfo {author} {\bibfnamefont {M.}~\bibnamefont {Xiao}},\ }\href@noop {}
  {\bibfield  {journal} {\bibinfo  {journal} {Advances in Optics and
  Photonics}\ }\textbf {\bibinfo {volume} {5}},\ \bibinfo {pages} {83}
  (\bibinfo {year} {2013})}\BibitemShut {NoStop}%
\bibitem [{\citenamefont {Berry}\ and\ \citenamefont {Klein}(1996)}]{BK96}%
  \BibitemOpen
  \bibfield  {author} {\bibinfo {author} {\bibfnamefont {M.~V.}\ \bibnamefont
  {Berry}}\ and\ \bibinfo {author} {\bibfnamefont {S.}~\bibnamefont {Klein}},\
  }\href@noop {} {\bibfield  {journal} {\bibinfo  {journal} {Journal of Modern
  Optics}\ }\textbf {\bibinfo {volume} {43}},\ \bibinfo {pages} {2139}
  (\bibinfo {year} {1996})}\BibitemShut {NoStop}%
\bibitem [{\citenamefont {Hannay}\ and\ \citenamefont {Berry}(1980)}]{HB80}%
  \BibitemOpen
  \bibfield  {author} {\bibinfo {author} {\bibfnamefont {J.~H.}\ \bibnamefont
  {Hannay}}\ and\ \bibinfo {author} {\bibfnamefont {M.~V.}\ \bibnamefont
  {Berry}},\ }\href@noop {} {\bibfield  {journal} {\bibinfo  {journal} {Physica
  D: Nonlinear Phenomena}\ }\textbf {\bibinfo {volume} {1}},\ \bibinfo {pages}
  {267} (\bibinfo {year} {1980})}\BibitemShut {NoStop}%
\bibitem [{\citenamefont {Matsutani}\ and\ \citenamefont
  {{\^O}nishi}(2003)}]{MO03}%
  \BibitemOpen
  \bibfield  {author} {\bibinfo {author} {\bibfnamefont {S.}~\bibnamefont
  {Matsutani}}\ and\ \bibinfo {author} {\bibfnamefont {Y.}~\bibnamefont
  {{\^O}nishi}},\ }\href@noop {} {\bibfield  {journal} {\bibinfo  {journal}
  {Foundations of Physics Letters}\ }\textbf {\bibinfo {volume} {16}},\
  \bibinfo {pages} {325} (\bibinfo {year} {2003})}\BibitemShut {NoStop}%
\bibitem [{\citenamefont {Romero-Cort{\'e}s}\ \emph {et~al.}(2016)\citenamefont
  {Romero-Cort{\'e}s}, \citenamefont {de~Chatellus},\ and\ \citenamefont
  {Aza{\~n}a}}]{LHA15}%
  \BibitemOpen
  \bibfield  {author} {\bibinfo {author} {\bibfnamefont {L.}~\bibnamefont
  {Romero-Cort{\'e}s}}, \bibinfo {author} {\bibfnamefont {H.~G.}\ \bibnamefont
  {de~Chatellus}}, \ and\ \bibinfo {author} {\bibfnamefont {J.}~\bibnamefont
  {Aza{\~n}a}},\ }\href@noop {} {\bibfield  {journal} {\bibinfo  {journal}
  {Optics Letters}\ }\textbf {\bibinfo {volume} {41}},\ \bibinfo {pages} {340}
  (\bibinfo {year} {2016})}\BibitemShut {NoStop}%
\bibitem [{\citenamefont {Luke}(1988)}]{luke1988sequences}%
  \BibitemOpen
  \bibfield  {author} {\bibinfo {author} {\bibfnamefont {H.~D.}\ \bibnamefont
  {Luke}},\ }\href@noop {} {\bibfield  {journal} {\bibinfo  {journal} {IEEE
  Transactions on Aerospace and Electronic Systems}\ }\textbf {\bibinfo
  {volume} {24}},\ \bibinfo {pages} {287} (\bibinfo {year} {1988})}\BibitemShut
  {NoStop}%
\bibitem [{\citenamefont {Chu}(1972)}]{chu1972polyphase}%
  \BibitemOpen
  \bibfield  {author} {\bibinfo {author} {\bibfnamefont {D.}~\bibnamefont
  {Chu}},\ }\href@noop {} {\bibfield  {journal} {\bibinfo  {journal} {IEEE
  Transactions on Information Theory}\ }\textbf {\bibinfo {volume} {18}},\
  \bibinfo {pages} {531} (\bibinfo {year} {1972})}\BibitemShut {NoStop}%
\bibitem [{\citenamefont {Lohmann}\ and\ \citenamefont
  {Thomas}(1990)}]{lohmann1990making}%
  \BibitemOpen
  \bibfield  {author} {\bibinfo {author} {\bibfnamefont {A.~W.}\ \bibnamefont
  {Lohmann}}\ and\ \bibinfo {author} {\bibfnamefont {J.~A.}\ \bibnamefont
  {Thomas}},\ }\href@noop {} {\bibfield  {journal} {\bibinfo  {journal}
  {Applied Optics}\ }\textbf {\bibinfo {volume} {29}},\ \bibinfo {pages} {4337}
  (\bibinfo {year} {1990})}\BibitemShut {NoStop}%
\bibitem [{\citenamefont {Leger}\ and\ \citenamefont
  {Swanson}(1990)}]{leger1990efficient}%
  \BibitemOpen
  \bibfield  {author} {\bibinfo {author} {\bibfnamefont {J.~R.}\ \bibnamefont
  {Leger}}\ and\ \bibinfo {author} {\bibfnamefont {G.~J.}\ \bibnamefont
  {Swanson}},\ }\href@noop {} {\bibfield  {journal} {\bibinfo  {journal}
  {Optics Letters}\ }\textbf {\bibinfo {volume} {15}},\ \bibinfo {pages} {288}
  (\bibinfo {year} {1990})}\BibitemShut {NoStop}%
\bibitem [{\citenamefont {Arriz{\'o}n}\ and\ \citenamefont
  {Ojeda-Castaneda}(1993)}]{arrizon1993talbot}%
  \BibitemOpen
  \bibfield  {author} {\bibinfo {author} {\bibfnamefont {V.}~\bibnamefont
  {Arriz{\'o}n}}\ and\ \bibinfo {author} {\bibfnamefont {J.}~\bibnamefont
  {Ojeda-Castaneda}},\ }\href@noop {} {\bibfield  {journal} {\bibinfo
  {journal} {Optics Letters}\ }\textbf {\bibinfo {volume} {18}},\ \bibinfo
  {pages} {1} (\bibinfo {year} {1993})}\BibitemShut {NoStop}%
\bibitem [{\citenamefont {Szwaykowski}\ and\ \citenamefont
  {Arriz{\'o}n}(1993)}]{szwaykowski1993talbot}%
  \BibitemOpen
  \bibfield  {author} {\bibinfo {author} {\bibfnamefont {P.}~\bibnamefont
  {Szwaykowski}}\ and\ \bibinfo {author} {\bibfnamefont {V.}~\bibnamefont
  {Arriz{\'o}n}},\ }\href@noop {} {\bibfield  {journal} {\bibinfo  {journal}
  {Applied Optics}\ }\textbf {\bibinfo {volume} {32}},\ \bibinfo {pages} {1109}
  (\bibinfo {year} {1993})}\BibitemShut {NoStop}%
\bibitem [{\citenamefont {Arriz{\'o}n}\ and\ \citenamefont
  {Ojeda-Castaneda}(1994)}]{arrizon1994multilevel}%
  \BibitemOpen
  \bibfield  {author} {\bibinfo {author} {\bibfnamefont {V.}~\bibnamefont
  {Arriz{\'o}n}}\ and\ \bibinfo {author} {\bibfnamefont {J.}~\bibnamefont
  {Ojeda-Castaneda}},\ }\href@noop {} {\bibfield  {journal} {\bibinfo
  {journal} {Applied Optics}\ }\textbf {\bibinfo {volume} {33}},\ \bibinfo
  {pages} {5925} (\bibinfo {year} {1994})}\BibitemShut {NoStop}%
\bibitem [{\citenamefont {Zhou}\ \emph {et~al.}(1999)\citenamefont {Zhou},
  \citenamefont {Stankovic},\ and\ \citenamefont {Tschudi}}]{zhou1999analytic}%
  \BibitemOpen
  \bibfield  {author} {\bibinfo {author} {\bibfnamefont {C.}~\bibnamefont
  {Zhou}}, \bibinfo {author} {\bibfnamefont {S.}~\bibnamefont {Stankovic}}, \
  and\ \bibinfo {author} {\bibfnamefont {T.}~\bibnamefont {Tschudi}},\
  }\href@noop {} {\bibfield  {journal} {\bibinfo  {journal} {Applied Optics}\
  }\textbf {\bibinfo {volume} {38}},\ \bibinfo {pages} {284} (\bibinfo {year}
  {1999})}\BibitemShut {NoStop}%
\bibitem [{Note1()}]{Note1}%
  \BibitemOpen
  \bibinfo {note} {The first row ($q$=2) in the table of \protect \citep
  {LHA15} is in error. It reads 1-1-1-1-1 but it should read
  1-3-1-3-1.}\BibitemShut {Stop}%
\bibitem [{\citenamefont {Maram}\ \emph {et~al.}(2014)\citenamefont {Maram},
  \citenamefont {Van~Howe}, \citenamefont {Li},\ and\ \citenamefont
  {Aza{\~n}a}}]{reza2014noiseless}%
  \BibitemOpen
  \bibfield  {author} {\bibinfo {author} {\bibfnamefont {R.}~\bibnamefont
  {Maram}}, \bibinfo {author} {\bibfnamefont {J.}~\bibnamefont {Van~Howe}},
  \bibinfo {author} {\bibfnamefont {M.}~\bibnamefont {Li}}, \ and\ \bibinfo
  {author} {\bibfnamefont {J.}~\bibnamefont {Aza{\~n}a}},\ }\href@noop {}
  {\bibfield  {journal} {\bibinfo  {journal} {Nature Communications}\ }\textbf
  {\bibinfo {volume} {5}},\ \bibinfo {pages} {5163} (\bibinfo {year}
  {2014})}\BibitemShut {NoStop}%
\bibitem [{\citenamefont {Hecke}(1981)}]{Hecke}%
  \BibitemOpen
  \bibfield  {author} {\bibinfo {author} {\bibfnamefont {E.}~\bibnamefont
  {Hecke}},\ }\href@noop {} {\emph {\bibinfo {title} {Lectures on the Theory of
  Algebraic Numbers}}},\ \bibinfo {edition} {2nd}\ ed.,\ \bibinfo {series}
  {Graduate Texts in Mathematics}, Vol.~\bibinfo {volume} {77}\ (\bibinfo
  {publisher} {Springer},\ \bibinfo {year} {1981})\BibitemShut {NoStop}%
\bibitem [{\citenamefont {Lang}(1986)}]{Lang}%
  \BibitemOpen
  \bibfield  {author} {\bibinfo {author} {\bibfnamefont {S.}~\bibnamefont
  {Lang}},\ }\href@noop {} {\emph {\bibinfo {title} {Algebraic Number
  Theory}}},\ \bibinfo {edition} {2nd}\ ed.,\ \bibinfo {series} {Graduate Texts
  in Mathematics}, Vol.\ \bibinfo {volume} {110}\ (\bibinfo  {publisher}
  {Springer},\ \bibinfo {year} {1986})\BibitemShut {NoStop}%
\bibitem [{\citenamefont {Ireland}\ and\ \citenamefont
  {Rosen}(1990)}]{IrelandRosen}%
  \BibitemOpen
  \bibfield  {author} {\bibinfo {author} {\bibfnamefont {K.}~\bibnamefont
  {Ireland}}\ and\ \bibinfo {author} {\bibfnamefont {M.}~\bibnamefont
  {Rosen}},\ }\href@noop {} {\emph {\bibinfo {title} {A Classical Introduction
  to Modern Number Theory}}},\ \bibinfo {edition} {2nd}\ ed.,\ \bibinfo
  {series} {Graduate Texts in Mathematics}, Vol.~\bibinfo {volume} {84}\
  (\bibinfo  {publisher} {Springer},\ \bibinfo {year} {1990})\BibitemShut
  {NoStop}%
\end{thebibliography}%

\end{document}